\documentclass[12pt]{article}
\usepackage[T1]{fontenc}
\usepackage{psfig}
\usepackage{bookman}
\usepackage{latexsym}
\usepackage{amssymb}
\usepackage{amsbsy}
\usepackage{cite}
\begin{document} 
  \title{The age problem and growing of structures for open system cosmology}
 \author{ M.de Campos$^{(1)}$ and  N.A.Tomimura$^{(1)}$ . }
\maketitle

   \footnote{
{ \small \it $^{(1)}$ Instituto de F\'\i sica \vspace{-0.2cm}}\\
{ \small \it Universidade Federal Fluminense  \vspace{-0.2cm}}\\
{\small \it Av. Gal. Milton Tavares de Souza s/n.$\!\!^\circ$, 
                  \vspace{-0.2cm} }}    
   \begin{abstract}
  A new equation for the density contrast is derived in the framework of reexamined
Newtonian cosmology taken into account adiabatic matter creation in the universe.

The age of the universe
and the reach of non linear regime of the density contrast  are usually
treated separately in the literature and this may lead to
controversial conclusions regarding the most adequate scenario  to describe
the universe. 

We relate the age of the universe and the growing mode of the density contrast
by introducing a variable $\xi $ that relates both of them, thus both aspects
are treated simultaneously.
We apply this procedure to the Friedmann type model where the
source of particle production is $\Psi = 3n\beta H $.
\end{abstract}
\section{Introduction}
In  1934, E.A.Milne analyzed the expansion of the universe using
elementary Newtonian theory \cite{Milne}.  In the same year W.H.Mc Crea and
E.A.Milne demonstrated the identity between the governing differential equations
for relativistic and Newtonian cosmology
\cite{Mac51}.
Three decades after, E.R.Harrison \cite{Har65} extended
Newtonian cosmology  to include pressure generalizing Mac Crea and Milne's
approach for dust matter.  The Harrison's approach is based on the
following set of equations:
\begin{eqnarray}
\frac{\partial \rho}{\partial t} +\vec{\nabla}_r \cdot (\rho +\frac{P_{th}}{c^2})
\vec{u} =0 \\
\frac{d \vec{u}}{d t} =-\vec{\nabla}_r \phi \nonumber  \\ 
\nabla ^2 _r \phi = 4\pi G (\rho + \frac{3P_{th}}{c^2})\, ,
\end{eqnarray}
which are respectively the continuity equation, Poisson's equation and the
motion equation.
Posteriorly, Harrison \cite{H} derived the motion equation by introducing a
non uniform pressure, namely:
\begin{equation}
\frac{\partial \vec{u}}{\partial t} +\vec{u} \cdot \vec{\nabla}_r \vec{u}
=-\vec{\nabla}_r \phi -(\rho +P_{th})^{-1} \nabla _r P_{th} \, .
\end{equation}
In the above equations $\rho,P ,u $ and  $ \phi$ denote respectively the
density of matter, the thermodynamic pressure, the velocity field and the gravitational potential.

The field equations for a homogeneous and isotropic universe are obtained by
using equations (1), (2) and (3).  These equations are identicals to the ones
obtained in the framework of general relativity in the comoving coordinate system.

However, it was observed in the end of the appendix II of the R.K.Sachs and
A.M Wolfe work that the perturbed modified field equations to the first order
are not the same in the general relativity and Newtonian cosmology.
Therefore, the density contrast equation for both theories are quite different.

This ambiguity was solved by Lima, Zanchin and Brandenberger \cite{Lima96} that
derived the  continuity equation in the modified
Newtonian approach to cosmology when pressure effects are included.
The new continuity equation is completely consistent with relativistic
perturbation theory and solves the contradiction  pointed out
by Sachs and Wolfe.

In this paper, we generalize the evolution equation for the density contrast obtained by Lima
et. al., for the universe with particle production and obtained the growing
and decaying modes for the density contrast.

In our work we introduce a variable ($\xi$) that depends of the age of the
universe and of the growing mode of the density contrast.  The use of $\xi $
can discern the cosmological model that furnishes a better
concordance with the experimental evidence about the age of the universe and
the time of structure formation

This paper is organized as follows:
we obtain in section II, in the linear approximation, the time evolution
equation for the mass density contrast for a Newtonian universe with matter
creation.  In section III we derive the time dependence of the density
contrast in which the source of particle production is $\Psi = 3n\beta H$.
We obtain a range for the parameter $\beta $ that updates the range obtained in
the literature  when the question of the age age of the universe
is involved. Finally we use the reach of the non linear regime to
establish a criterion about the most adequate cosmological scenario.
%
%
%
%
%
%
%
%
%
\section{ Equation for the density contrast}
In the context of matter creation the thermodyamical conservation law reads:
\begin{equation}
      \frac{d \rho }{dV}  + \frac{1}{V} (P_{th} + \frac{\rho }{c^2} ) + \frac{h}{nV} \frac{d}{dV} (nV) = 0 \, ,
        \end{equation}
where $n $ is the particle number density, $h = \rho +P_{th} $ is the enthalpy per
unity volume and  $P_{th} $ is the thermodynamical pressure.  Consequently the
continuity equation becomes
\begin{equation}
\frac{\partial\rho}{\partial t} + \nabla_{r} \cdot (\rho \vec {u} ) + \frac{P_{th}}{c^{2}}
                     \nabla _{r} \cdot \vec {u}= \frac{h}{n} \Psi ,
\end{equation}
where $\Psi $ is the source of particle production.  The other hydro-dynamical
equations that describe the cosmic fluid are the momentum conservation
equation and Poisson's equation, respectively:
 \begin{equation}
\frac{\partial \vec{u}}{\partial t}+\vec{u} \nabla _{r}\vec{u}=-\nabla _{r}\Phi -(\rho +\frac{P}{c^{2}})^{-1}\nabla _{r}P \, ,
     \end{equation}
and
\begin{equation}
\nabla_{r}^{2}\Phi=4\pi G( \rho+\frac{3P}{c^{2}}) \, ,
\end{equation}
$\Phi $ is the gravitational potential and $P $ is the total pressure that
includes the thermodynamical pressure and the creation pressure $\tilde{P}$
, so
\begin{equation}
P = \tilde{P} + P_{th}\, .
\end{equation}
The creation process will be considered adiabatic, consequently the pressure
creation is given by \cite{Calvao} :
\begin{equation}
\tilde{P} = - \frac{h \Psi }{n \theta } \, .
\end{equation}
In order to obtain a cosmological scenario with matter creation one still
needs to provide the state equation 
\begin{equation}
P_{th} = \nu \rho \, .
\end{equation}

Using the field equations for FRW model \cite{Lima96b} with particle
production and the state equation above, we can write a differential equation
for the scale factor:
\begin{equation}
R\ddot{R}+(\frac{3\nu +1}{2}-\frac{(\nu+1)\Psi}{2nH})\dot{R}^2 + (\frac{3\nu
  +1}{2}-\frac{(\nu+1)\Psi}{2nH})\kappa =0
\end{equation}
 
Our main aim in this section is to find the differential equation that governs
the evolution of the
density contrast in a Universe where the particle production is given by $\Psi
= 3n\beta H $, where $\beta $ is a constant.  Note that $\beta = 0 $ 
implies $\Psi = 0 $ and we recover the standard Friedmann model.

To study the evolution of small fluctuations in an expanding universe we
consider the standard perturbation ansatz:
\begin{equation}
           \rho = \rho _{b}(t) \{1+\delta (r,t) \} \, ,
               \end{equation}
           \begin{equation}
             P_{th}=P_{thb}(t)+\delta P_{th}(r,t) \, ,     
                \end{equation}
              \begin{equation}
                     \Phi = \Phi _{b} (r,t)+ \varphi (r,t) \, , 
                             \end{equation}
                     \begin{equation}
                         \vec{u} = \vec{u_{b}} (t) + \vec{v}(r,t) \, , 
                                \end{equation}
                     \begin{equation}
                             \tilde{P} =\tilde{P_{b}} (t)+\delta \tilde{P} (r,t) \, .
                                      \end{equation}
The quantities carrying the subscript {\it b } represent the homogeneous solution to
the unperturbed equations.
Inserting the above expressions into eqs.(2),(3) and (4) we get to first order in
perturbations
\begin{equation}
       \dot{v} + \frac{\dot{R}}{R} v=\frac{-1}{R} \nabla \varphi - \frac{\nu - \beta (\nu + 1)}{(\nu + 1)(1-\beta)}
           \frac{\nabla \delta}{R} 
           \end{equation} 
               
            \begin{equation}
                     \nabla^{2}\varphi = 4\pi G \rho_{b}(t)R^{2}\delta \{ 1+3\nu - 3(\nu +1)\beta \} 
                             \end{equation}
      \begin{equation}                           
         \dot{\delta} = \frac{\nu+1}{R}\nabla v (1-\beta) \, .
              \end{equation}
Change to comoving coordinates following standard lines \cite{Pee80}, eliminating the
peculiar velocity from equations (17) and (19) and using (18), is ready that
\begin{eqnarray}
             & \ddot{\delta }&+2\frac{\dot{R}}{R}\dot{\delta }+4\pi G\rho_{b}(t)\delta \{ 1+3\nu-3(\nu +1)\beta \} 
                     (\nu + 1)(\beta -1)  \nonumber \\
&=&(\nu -\beta (\nu + 1))\frac{\nabla ^{2} \delta }{R^{2}} \, .
                         \end{eqnarray}
This is the differential equation that governs the evolution of density contrast
in the presence of matter creation, when we describe a matter distribution
with uniform pressure using the modified Newtonian equations.
It is spatially homogeneous, so we expect to find plane wave solutions.
The difficulty to apply Newtonian cosmology in the study of
scalar perturbations is related with the impossibility of the Newtonian
scenario to describe long wavelengths perturbations.
The cosmological perturbations can be of the two kind  according to the
wavelength $\lambda$, so that $\lambda > d_H $ or $\lambda < d_H $, 
where $d_h $ is the Hubble sphere.
Perturbations with wavelengths larger than Hubble scale requires some form of
a general relativistic theory of perturbations. 
For wavelengths smaller than Hubble scale  the evolution of mass density
 can be studied using Newtonian
theory.  Application of Newtonian equations is further restricted to
non-relativistic matter and cannot be used for relativistic components even
for scales smaller than Hubble radius.

The reexamined Newtonian equations  indicate a way of obtaining
the same evolution equation for the density contrast as could be by the full
relativistic approach.  In this way one can extend the domain of validity of
equation (20) in order to analyze perturbations  even in the radiation
dominated phase and to apply the
large  wavelength limit.

Next, we derive the solution of the density contrast when the source of
particle production is $\Psi = 3n\beta H $, concentrating our attention on the growing mode.
The principal advantage of this approach is the relative facility to obtain the
perturbed equations for the density contrast which is the same equation as in
general relativity when $\beta = 0 $ in the synchronous and comoving
coordinates.

Although the gauge invariant approach is conceptually more attractive since there is no
need to identify the physical  and unrealistic space-times, nevertheless it is more
complicated and the physical meaning of gauge independent variables do not in
general possess any simple interpretation.
Besides, the gauge mode that emerges from the differential equation for the
density contrast is generally related with some kind of decaying mode when
the universe is homogeneous and isotropic \cite{Hwang}.

%
%
%
%
%
%
\section{Choosing a model}
If the space time has uniform curvature, the line element is given by
\begin{equation}
              ds^{2}=dt^{2}-R^{2}(t)(\frac{dr^{2}}{1-kr^{2}}+r^{2}d \theta ^{2}+r^{2}sin^{2}(\theta)d \phi ^{2}) \, .
                      \end{equation}
The evolution of the scale factor  is obtained from solutions of differential
equation \cite{Lima96b}
\begin{equation}
              R\ddot{R}+\Delta (k + \dot{R^{2} ) }=0 \, ,
               \end{equation}
where
     \begin{equation}
          \Delta = \frac{3}{2} (\nu +1)(1- \beta )-1 \, .
                     \end{equation}
Explicit solutions for differential equation (22) when $\kappa = \pm 1$ for
any value of $\beta $ are not easy to obtain.  We have managed to find a class of exact
solutions for the following cases:
\begin{equation}
\kappa = 0 \, \, \beta \neq 1  \longrightarrow \,  R = R_{0}(\frac{t}{t_{0}})^{\frac{1}{\Delta +1}}
 \end{equation} \, , \\
\begin{equation}
 \kappa = 0 \, \, \beta = 1 \longrightarrow \, R = R_{0}e^{\frac{t}{t_{0}}} \, ,
 \end{equation}\\
\begin{equation}
\kappa = \pm 1 \, \, \beta = 1 \longrightarrow \, R = \frac{1}{2}\{\frac{\kappa + e^{\pm
                          2C_{1}^{\frac{1}{2}}}(t+C_{2})}{e^{\pm C_{1}^{\frac{1}{2}}}(t+C_{2})
                                C_{1}^{\frac{1}{2}}} \} \, .
 \end{equation}\\
The subscript $ 0 $ alludes to the present time and $C_{1} $ and $C_{2} $ are integration constants. 

Whether the Universe turns out to be spatially closed, open or precisely
flat remains an empirical question.
The recent results of the Boomerang provide convincing evidence in favor of
the standard paradigm: The Universe is flat, however the possibility that the
universe might be spatially closed do not be discarded
\cite{Martin}. Consequently, we consider the Universe as flat for
subsequent calculation.

Using the scale factor (24) then the mass-energy density is given by:
\begin{equation}
 8\pi G \rho_{b}  = \frac{1}{(\Delta + 1)^2 t^2 }\, .
       \end{equation}
Using equations (27), (24) and the long wavelength limit into equation (20) we finally obtain
\begin{equation}
               \ddot{\delta}+\frac{\alpha}{t} \dot{\delta}+(\alpha -2)
               \frac{\delta}{t^{2}} = 0 \, ,
                    \end{equation}
where
         \begin{equation}
             \alpha=\frac{4}{3(\nu +1) (1-\beta)} \, .
                      \end{equation}
                  
Solutions  for  equation (28) are 
\begin{eqnarray}
\delta_{-} =  C_{1} t^{-1} \, , \\
\delta_{+} = C_{2}t^{2-\frac{4}{3(1-\beta)(\nu +1)}} \, .
\end{eqnarray}
In the absence of particle production, $\beta = 0 $, relations (30) and (31) become the
usual density contrast for Friedmann models without creation \cite{Wei72}.
Although the decreasing mode can be important in some circumstances,  the
increasing mode is responsible for the formation of cosmic structures in the
gravitational instability picture. Taking
into account the decreasing mode the universe would not have been homogeneous in
the past.  Besides, Peebles argued that a growing mode that starts to grow just
after the end of radiation era has a negligible component of decaying solution
 \cite{Peb}.

One of the problems of the standard FRW model is related to the age of the
universe; it is younger than it should be according to the experimental values.
However, in the framework of particle production the value of the age of the
universe is more compatible with the experimental evidences.  To show
this we define  the quotient $ \zeta $ given by:
\begin{equation}
\zeta = \frac{H_c}{H_F} \, ,
\end{equation}
at the same time coordinate. The subscript F and C refer respectively to the
Friedmann standard model and the model with particle creation.

Considering the scale factor (24) one obtains
\begin{equation}
\zeta = \frac{1}{1-\beta} \, .
\end{equation}
 If we expect that the open system model furnishes an oldest universe, then
 $\zeta > 1$ implying $ 0< \beta <1$.
Using the oldest globular clusters, B. Chaboyer \cite{Cha} infers a range for
 the age of the universe, namely  $9.6 Gyr < t_{universe} < 15.4 Gyr$.
Consequently, the  Hubble constant for the standard model lies in
 the interval
\begin{equation}
43.3 < H_0 < 69.4 \, . 
\end{equation}
This range does not agree with the interval for $H_0$ estimated recently by
Willick and Batra \cite{Batra}, namely:
\begin{equation}
80 < H_0 < 90 \, . 
\end{equation}
One can easily verify that there is a small overlap between (34) and
the values obtained by Richtler and Drenkhahn \cite{Ton},
given by
\begin{equation}
68 < H_0 < 76 \, . 
\end{equation}

Using the scale factor (24) for dust matter universe, the Hubble function for
a universe with particle production is
given by
\begin{equation}
  H = \frac{2}{3(1- \beta )t}  \nonumber . 
\end{equation} 
We can conciliate the ranges (35) and (36) using the Hubble function above, where the creation parameter $\beta $ lies in the interval
\begin{equation}
0.13 < \beta < 0.52 \, . 
\end{equation}
This result updates the interval for $\beta$ determined by Lima et
al \cite{Lima96b} ($0.34 < \beta < 0.60 $).

Now, we study the reach of non linear regime, identifying the epoch for the
reach of the non linear regime with the epoch for the formation of the super
clusters. In addition, we suppose that the star formation is posterior to the
formation of the super clusters.  First, because  the density of the super cluster in the universe is
approximately equal to the universe density.  Second, considering the
universe to form ``from top down'' is acceptable for some type of models, for
example the hot dark matter model \cite{Kolb}.  In other words, we want to infer
that the star formation is posterior to the super cluster formation and,
consequently, to the reach of the non linear regime.
The growing mode for the density contrast for the FRW standard model can be
written as
\begin{equation}
\delta_+ = \delta_d \{\frac{t}{t_d}\} ^{2/3} \, ,
\end{equation}
where the subscript $d$  refers to decoupling time.
Typically we have $(\frac{\delta \rho}{\rho})_{dec} < (10^{-2} - 10^{-3})$
\cite{Kolb}, for a decoupling time of the order $10^5 ys$ .  Substituting the
decoupling time and the corresponding anisotropy, the reach for the non linear
regime for FRW standard model occurs in the range $10^{-1} Gyr < t < 3.16 \times
Gyr $.  Note that, the standard model furnishes a sub estimate age for
the universe, around 8 Gyr.  Now, if we consider the difference between the age of
the universe and the reach of the non-linear regime we obtain the interval $5 Gyr
< t < 7 Gyr$ for the age of the super cluster. In the literature it is known that the age for the oldest stars
in our galaxy is around 12 Gyr \cite{Cha}.  

The open system cosmology furnishes a reasonable solutions for the age
of the universe.  What
happens with the reach of the non linear regime for the density contrast?

To solve this questions we introduce a variable $\xi$ that relates the age of
the universe and the growing mode for the density contrast in order to
discern among more suitable cosmological model, it is given by:
\begin{equation}
\xi = \frac{\delta_{+c}}{\delta_{+F}} \zeta \, .
\end{equation}

 A  cosmological model with better
characteristics to represent the universe must be satisfied if $\xi > 1$.
Taking into account the growing mode (31) and relation (33), the condition (40)
implies that $\beta < 0$. Negative values for the parameter $\beta$ violates
the second law of thermodynamics.  Then we conclude that the respective cosmological
model to the source given by $\Psi = 3n\beta H $ does not give a better result than the usual FRW model
when the age problem and the reach of the non-linear regime are taking
together.
On the another hand if we examine the difference of the age for the universe and
the reach for the non linear regime in the context of the open system
cosmology with a particle source given by $\Psi = 3n\beta H$, we note that
this difference is smallest than the result obtained in the FRW framework.

\end{document}